\documentclass[aps,pre,10pt,superscriptaddress,showpacs,showkeys,twocolumn]{revtex4}
\usepackage{graphicx}
\usepackage{amssymb}        
\usepackage{xcolor}

 \providecommand{\un}[1]{\mathrm{#1}}
 \providecommand{\acknowledgments}{\section*{Acknowledgments}}
\newcommand{\clabel}[1]{\label{#1}} 
\newcommand{\w}{\omega}
\newcommand{\Ah}{\frac{A}{2}} 
\providecommand{\dfrac}[2]{\frac{\displaystyle #1}{\displaystyle
#2}}
 \newcommand{\Kt}{K$^+$\,}  
\newcommand{\cs}{c_s}
\begin{document}
\title{Ultradiscrete kinks with supersonic speed in a layered crystal with realistic potentials}
\author{J. F. R. Archilla}
\email[]{archilla@us.es}
\affiliation{Grupo de F\'{\i}sica No Lineal,  Universidad de
Sevilla, ETSI Inform\'{a}tica,
  Avda Reina Mercedes s/n, 41012-Sevilla, Spain}
\author{Yu. A. Kosevich}
\email[Corresponding author:  ]{yukosevich@gmail.com}
\affiliation{Semenov Institute of Chemical Physics, Russian
Academy of Sciences.  Kosygin street 4, 119991 Moscow, Russia}
\author{N. Jim\'enez}
\affiliation{Instituto de Investigaci\'{o}n para la Gesti\'{o}n,
Integrada de las Zonas Costeras, Universidad Polit\'{e}cnica de
Valencia, C/.Paranimfo 1,  46730 Grao de Gandia, Spain}
\author{V. J. S\'{a}nchez-Morcillo}
\affiliation{Instituto de Investigaci\'{o}n para la Gesti\'{o}n,
Integrada de las Zonas Costeras, Universidad Polit\'{e}cnica de
Valencia, C/.Paranimfo 1,  46730 Grao de Gandia, Spain}
\author{L. M. Garc\'{i}a-Raffi}
\affiliation{Instituto Universitario de Matem\'{a}tica Pura y
Aplicada, Universidad Polit\'{e}cnica de Valencia, Camino de Vera
s/n, 46022 Valencia, Spain }%

\newcommand{\K}{K^+ }
\newcommand{\kc}{\mathrm{k}_\mathrm{e}}
\newcommand{\e}{\mathrm{e}}


\begin{abstract}
In this paper we develop a dynamical model of the propagating
nonlinear localized excitations, supersonic kinks, in the cation
layer in a silicate mica crystal. We start from purely
electrostatic Coulomb interaction and add the
Ziegler-Biersack-Littmark short-range repulsive potential and the
periodic potential produced by other atoms of the lattice. The
proposed approach allows the construction of supersonic kinks
which can propagate in the lattice within a large range of
energies and velocities. Due to the presence of the short-range
repulsive component in the potential, the interparticle distances
in the lattice kinks with high energy are limited by physically
reasonable values. The introduction of the periodic lattice
potential results in  the important feature that the kinks
propagate with the single velocity and single energy which are
independent on the excitation conditions. The unique average velocity of
the supersonic kinks on the periodic substrate potential we relate
with the kink amplitude of the relative particle displacements
which is determined by the interatomic distance corresponding to
the minimum of the total, interparticle plus substrate, lattice
potential. The found kinks are ultra discrete and can be described
with the "magic wave number" $q\simeq 2\pi/3a$, which was
previously revealed in the nonlinear sinusoidal waves and
supersonic kinks in the Fermi-Pasta-Ulam lattice.  The extreme
discreteness of the observed supersonic kinks,  with basically two
particles moving at the same time, allows the detailed
interpretation of their double-kink structure which is not
possible for the multi-kinks without an account for the lattice
discreteness. Analytical calculations of the displacement patterns
and energies of the supersonic kinks are confirmed by numerical
simulations. The computed energy of the found supersonic kinks in
the considered realistic lattice potential is in a good agreement
with the experimental evidence for the transport of localized
energetic excitations in silicate mica crystals between the points
of $^{40}$K recoil and subsequent sputtering.
\end{abstract}
\maketitle

\section{Introduction}
Many minerals are known by their capability of recording the
tracks of charged particles and are often used as solid state
nuclear track detectors (SSNTDs)\cite{durrani2001,durrani2008}.
Among them, silicate mica muscovite crystal has been relevant as
it was the second material and the first natural one where the tracks
from fission fragments were found~\cite{silkbarnes59}. Soon
later fossil tracks were also found in silicate mica
crystal~\cite{pricewalker62}. This crystal  seems to be one of the
most sensitive of the natural SSNTD~\cite{fleischer2011}.  Mica
crystal has important technological applications due to its
dielectric and heat shielding properties. Due to its heat
resistance, mica crystal can be used inside a nuclear reactor core
for particle detection. It has also been used in geochronology and
to probe the existence of dark matter~\cite{snowdenifft1995}, to
find exotic nuclear reactions, decays of super heavy elements and
weakly interacting massive
particles~\cite{durrani2001,durrani2008}.
Silicate mica crystals are among the materials considered as engineered
barriers for nuclear waste storage due to their high rate of
reaction with heavy ions in low temperature reconstructive
transformations~\cite{AmMin01a}. The latter are also
important  for the development of advanced structural
ceramics~\cite{HYISNM02}. The reconstructive transformations at low temperature
were related with the existence of $stationary$ intrinsic localized vibrational modes
(discrete breathers) in mica crystals~\cite{ACANT06,DSA11}.

\begin{figure}[t]
\centerline{\includegraphics[width=0.9\columnwidth]{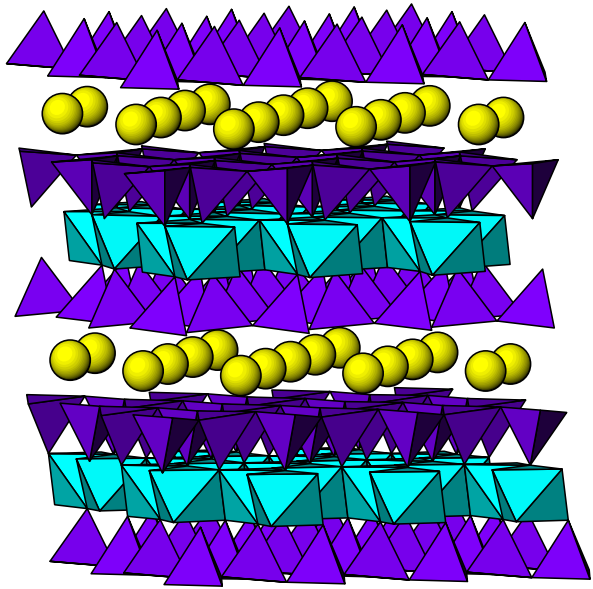}}
\caption{Representation of the mica structure from the point of
view that emphasizes the close-packed lines of the $\K$ hexagonal
layer represented by yellow balls. For a view from the top, see
Fig.~\ref{fig:micahexagonal}} \clabel{fig:micarows}
\end{figure}

Tracks of positrons, muons and other particles have been
reported~\cite{russell67b,russell68,russell88a,russell88b} in mica
muscovite crystals.
 Some of these tracks were identified as being produced by
positrons resulting from the $\beta +$ decay of $^{40}$K. This
isotope is relatively abundant in the minerals and can also
experience the $\beta-$, electron capture and other kinds of
decays~\cite{NDSA40,radionuclides2012}.
Most of the tracks, however, cannot be explained as being produced
by charged particles but could have been produced by some kind of
$propagating$ energetic vibrational excitation since the tracks
are along the close-packed lines of the $\K$ hexagonal layer shown
in Figs.~\ref{fig:micarows} and \ref{fig:micachain}.  One
interpretation of these tracks is that they are formed by the
quasi-one-dimensional (1D) localized nonlinear excitations,
sometimes called {\em quodons}~\cite{schlosserrussell94}, whose
exact nature is still unknown.

A likely source for the vibrational energy required to initiate a
quodon is the recoil energy of the $^{40}$K after $\beta$ decay, which can be
up to 52~eV~\cite{NDSA40,radionuclides2012}.
In mica muscovite crystals, there are about three decays per
second and cm$^3$, so after many years of the sensitive period,
when track recording is possible, there are many possibilities to
initiate a quodon. An experiment was done to shed some light into
the relationship  between tracks and
quodons~\cite{russelleilbeck07}. A mica specimen was irradiated by
alpha-particles and the ejection of atoms was detected on the
opposite side, along the close-packed lines. The ejected atoms
could not been identified and sputtering energies are not known
exactly; however from experimental and theoretical studies, they
are known to be in the range of $4$-$8$~eV~\cite{kudriavtsev2005}.

\begin{figure}[h]
\centerline{\includegraphics[width=0.9\columnwidth]{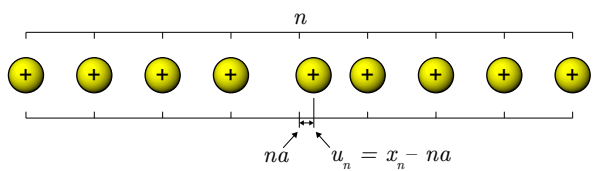}}
\caption{Representation of a close-packed line of  $\K$ ions and
the variables used. At equilibrium one has $x_n=n a$, and $u_n$
measures the displacement from equilibrium of the ion labeled $n$.
In physical units $a=5.19$~\AA\mbox{}, and $a=1$ in scaled units.}
\clabel{fig:micachain}
\end{figure}

In an attempt to understand this phenomenon, numerical simulations
were conducted in an idealized 1D model where the particles with a
given high energy hit the boundary of the lattice \cite{Dou11}.
This study reports the formation of breathers and kinks in the
lattice bulk and the ejection of atoms at the opposite boundary of
the sample. However, the model did not intend to use realistic
values of the mica muscovite parameters.

A minimal model of the cation lattice with realistic parameters
has been proposed recently~\cite{archilla2013,archilla2014medium}. Only
\Kt ions were considered with the  actual potassium mass and
Coulomb interaction between them. The interaction with the rest of
the lattice was implicitly considered as the force keeping the
ions inside the crystal with the known interatomic distance in the
muscovite. In those publications, it was reported the existence of
supersonic kinks which were easily generated within a wide range
of velocities and energies.

However such model is too simplistic because for relatively high
ion energies it results in unrealistically small distances between
the ions, of tenths of angstrom. In this paper, we develop a
comprehensive dynamical model of the $propagating$ nonlinear
localized lattice excitations, {\it supersonic kinks}, in which we
use the Ziegler-Biersack-Littmark (ZBL) short-range repulsive
potential, introduced in particle bombardment
studies~\cite{ziegler2008}. ZBL potential describes the Coulomb
repulsion between nuclei, which is partially screened by the atom
electrons and rapidly decays within few \AA.  The results for the
moving energy-carrying objects in such potential  are similar to
those obtained with pure Coulomb
potential~\cite{archilla2013,archilla2014medium}, namely that
supersonic kinks, traveling without attenuation for long
distances, can be produced within a wide range of velocities and
energies. The interparticle distances in the kinks in our
dynamical model are limited by physically reasonable values. Note
that our dynamical  model allows for bond dissociation, which realizes one possible physical mechanism of the normal
heat transport in low-dimensional
systems~\cite{savin-kosevich2014}. From the other hand, the classical description of the supersonic kink formation and propagation is justified by the fact
that ZBL potentials are known to provide very good agreement with the
experiments on atom collisions in the context of radiation damage and ion track formation, which can be modeled using the methods of classical molecular dynamics \cite{gibson1960,meftah1994,trautmann2000}.

The literature about kink propagation in lattices with different
inter-particle and on-site potentials is extensive. The most
studied and generic model is the Frenkel-Kontorova (FK) model
\cite{FKM}, see reviews in Refs.~\cite{chaikin95,BK98,braun2004}.
However, most of the kinks considered in these publications are
subsonic ones. Supersonic kinks in the systems with substrate have
been found in models with anharmonic interparticle
coupling~\cite{kosevich73,milchev90,savin95,zolotaryuk97}. They
have the property than only a discrete set of velocities allows
the propagation of the kinks without attenuation. They can be
described as multi-kinks or lattice $N$-solitons depending on
whether the description is done in terms of displacements, strains
or velocities.

The FK model has also been considered in layered materials, to
model, for example, the in-plane dynamics of a few-layer graphene,
in order to explain the results of molecular dynamics simulation
of the cross-plane thermal conductance~\cite{ni-kosevich2014}.

As a next step towards a more realistic description, we construct
explicitly the interaction with the surrounding atoms, using
standard empiric potentials introduced in molecular dynamics,
which give rise to a periodic non-sinusoidal substrate potential.
Supersonic lattice kinks, which are also called {\em
crowdions}~\cite{kosevich73}, were constructed in our dynamical
model. They propagate in such potential with a velocity which is
independent of the excitation conditions and is determined only by
the lattice potential parameters.
The kink (crowdion) with the unique supersonic velocity,
propagating in our dynamical model of the cation layer on a
substrate, has a double-kink structure. We relate the unique
velocity of the supersonic kink on the periodic substrate
potential with the kink amplitude of the relative particle
displacements which is determined by the interatomic distance
corresponding to the $minimum$ of the {\it total}, interparticle
plus substrate, lattice potential.

The important characteristic, found in the kinks studied in the
present work, is extreme discreteness of the kinks, with only one
or two particles in motion at a given time. The extreme kink
discreteness allows the complete understanding in physical terms
of the mechanism that brings about the double-kink structure of
the found kinks.

The computed energy of the found supersonic kinks in the
considered realistic lattice potential is approximately 26~eV:
Such energy can be provided by the recoil of isotopes of potassium
after radioactive decay and it is larger than the sputtering
energy. This value of the characteristic energy of the found
supersonic kinks allows to assume that  the tracks found in mica
muscovite crystals can be related with the propagating lattice
kinks (crowdions).

We describe the Hamiltonian systems, in which the supersonic kinks
under certain conditions lose their energy by emitting phonons;
see Sec.~V below. But nevertheless the considered systems continue
to be the strongly $underdamped$ systems. Therefore the known
properties of the kinks (fronts) in $dissipative$ $overdamped$
systems \cite{pomeau1986,cross1999}, including discrete ones
\cite{clerc2011,matthews2011}, cannot be applied directly to the
considered ultra-discrete supersonic kinks. For instance, due to
the kink discreteness on the atomic scale, the position of the
kink (front) core cannot be unambiguously defined as the position
of the maximum of the displacement derivative $\partial_x u_x$
\cite{clerc2011} because the latter quantity is not well defined
for the ultra-discrete kink. Therefore we can define and
numerically measure only the $average$ kink velocity and cannot
measure the possible weak time-domain oscillations of the kink
speed and snaking bifurcation diagram
\cite{clerc2011,matthews2011}.

The paper is organized as follows. First we review and extend the
results obtained with only Coulomb interaction, using the
sinusoidal waveform proposed for supersonic kinks in the
Fermi-Pasta-Ulam (FPU)
chain~\cite{kosevich93,kosevich04,poggi1997}. The sinusoidal
waveform is a good description for $\lambda\simeq 3a$, where
$\lambda$ is a characteristic wavelength of the sinusoidal
waveform and $a$ is the lattice constant, but fails for
$\lambda\simeq 2a$, been replaced by an almost triangular
waveform, corresponding to nearly hard-sphere collisions.
Afterwards, we consider the effects of long-range interactions
with several neighbors, and introduce a short-range
nearest-neighbor ZBL potential. Thereafter, the substrate
potential is constructed and the properties of the single-velocity
lattice kinks in the cation layer on the substrate are analyzed
with detail. Finally we provide a summary and discussion of all
the main results of the paper.

\section{Model and sinusoidal kink description}
\clabel{sec:model}

We consider as a starting point a 1D model for the dynamics of
$\K$ ions, given in dimensionless form by
\begin{eqnarray}
\ddot  u_n=-\frac{1}{(1+u_{n+1}-u_{n})^2}+\frac{1}{(1+u_{n}-u_{n-1})^2} .
\clabel{eq:forcesadim}
\end{eqnarray}
which describes a chain of ions coupled to their nearest neighbors
by electrostatic potential. The variables $u_n$ represent the
displacement of a particle in the chain with respect to its
equilibrium position, normalized to the lattice constant. The
values of the scaled units are the following: for lengths, the
lattice constant, the equilibrium distance between $\K$ ions,
$u_L=a=5.19$~\AA; for masses, the mass of a $\K$ ion,
$m_{\K}=39.1$~amu; for time, $\tau=\sqrt{m_{K} a^3/ \kc\,
\e^2}\simeq 0.2$~ps, $\kc$ is the Coulomb constant and $e$ is the
elementary unit of charge. Other physical units in the system are
velocity $u_V=2.6~\un{km/s}$, energy $u_E=2.77\,\un{eV}$, momentum
$u_P=1.698\times 10^{-22}$\,kg\,m/s and frequency 5~THz. The
dimensionless speed of sound in this system is $\cs=\sqrt{2}$, or
3.7~km/s in physical units.

For small amplitudes, the  potentials in Eq.~(\ref{eq:forcesadim})
can be expanded in series using that $1/(1+y)^2\simeq
1-2y+3y^2-4\,y^3\dots $. Retaining cubic and smaller terms, we
obtain:
\begin{eqnarray}
\ddot u_n=&\mbox{}&\cs^2\big[(u_{n+1}+u_{n-1}-2 u_{n})\quad
\nonumber\\
\mbox{}&-&3/2(u_{n+1}-u_{n})^2+3/2(u_{n}-u_{n-1})^2\quad \nonumber \\
\mbox{}&+&2(u_{n+1}-u_{n})^3-2(u_{n}-u_{n-1})^3+\dots  \big] \, ,
\label{eq:alfabetafpu}
\end{eqnarray}
which are the $\alpha$-$\beta$ FPU equations of motion. We would
like to emphasize that the $\alpha$-$\beta$ FPU equation
(\ref{eq:alfabetafpu}) describes the Coulomb lattice
(\ref{eq:forcesadim}) only in the small and intermediate-amplitude
limit and is not applicable to this lattice in the large-amplitude
limit, see Fig.~4 and the subsection dedicated to the triangular
waveform below.

 Without nonlinear
terms, Eq.~(\ref{eq:alfabetafpu}) is reduced to the discrete linear wave equation:
\begin{eqnarray}
\ddot u_n=\cs^2 (u_{n+1}+u_{n-1}-2 u_{n}) \, , \label{eq:linear}
\end{eqnarray}
where $\cs$,  the speed of sound, is the long wavelength phonon
velocity. It is also both the maximum phase and group velocity.
Note that in our scaling $\cs=\sqrt{2}$.

Introducing a new variable, the deformation from the equilibrium
position or {\em strain} $v_n=u_n-u_{n-1}$, the equations above
can be written as:
\begin{equation}
\ddot{v}_n=2 F_n-F_{n+1}-F_{n-1} \, , \quad \mathrm{with}\quad
F_n=\frac{1}{(1+v_n)^2} \, , \label{eq:ddotvn}
\end{equation}
where $v_n=0$ corresponds to the unperturbed lattice. The boundary
conditions assume fixed particles at the ends of the lattice. Kinks
are produced numerically by applying at the chain boundary a {\em
half-wave} perturbation, which is a sinusoidal displacement during
half of a period, starting and finishing at the equilibrium
position~\cite{archilla2013,kosevich04}. In order to describe  a
kink traveling to the right, the
following {\em ansatz} was introduced in Refs.~\cite{kosevich93,kosevich04}:
\begin{equation}
v_n=-\frac{A}{2}(1+ \cos(q n-\omega t))\quad \mathrm{if}\quad -\pi
\leq q n-\omega t<\pi \, , \clabel{eq:vn}
\end{equation}
and $v_n=0$ otherwise, where $A$ is the kink amplitude of the $relative$ particle displacements. The bonds
are always compressed so $v_n$ is negative with a minimum value of
$-A$,  corresponding to the maximum compression of the bond. For
analogy,  we use the usual wave terminology, so $\phi_n=\omega
t-q\,n$ is the phase; $q$ is the effective wave number; $\omega$ is the angular
frequency; $T$ is the period; $\lambda=2\pi/q$ is the wavelength.
The $average$ $velocity$ of the kink is defined as $V=\omega/q$: It is determined by the time the kink needs to travel a lattice
site.
Numerically it is measured by tracing  the states of maximum
lattice compression.

The sinusoidal waveform with the $magic$ $wave$ $number$
$q=2\pi/3$ is an exact solution for the nonlinear sinusoidal waves
in the FPU chain \cite{kosevich93,kosevich04,poggi1997}, which is
characterized by the translational-invariant interparticle
potential with cubic and quartic anharmonicity. This potential is
a good approximation for many realistic potentials, including the
Coulomb one, for the intermediate amplitudes. Such waveform
provides a very useful framework for understanding the relative
phases of the moving particles and the behavior of the kink.

Figure~\ref{fig:profiles}~(top) compares the result of the
numerical simulation for displacements and strains, with the
corresponding analytical expressions derived from
Eq.~(\ref{eq:vn}), for the  intermediate value of the amplitude
$A=0.55$ and $magic$ $wave$ $number$ $q\simeq 2\pi/3$. The analytical
expression fits very well the numerical data.
Figure~\ref{fig:profiles}~(bottom) pictures  a kink with
wavenumber $q=\pi$. There are not enough particles to compare with
the analytical form, but the main properties of the latter are
valid.

\begin{figure}[h] \begin{center}
\centerline{\includegraphics[width=8cm,clip]{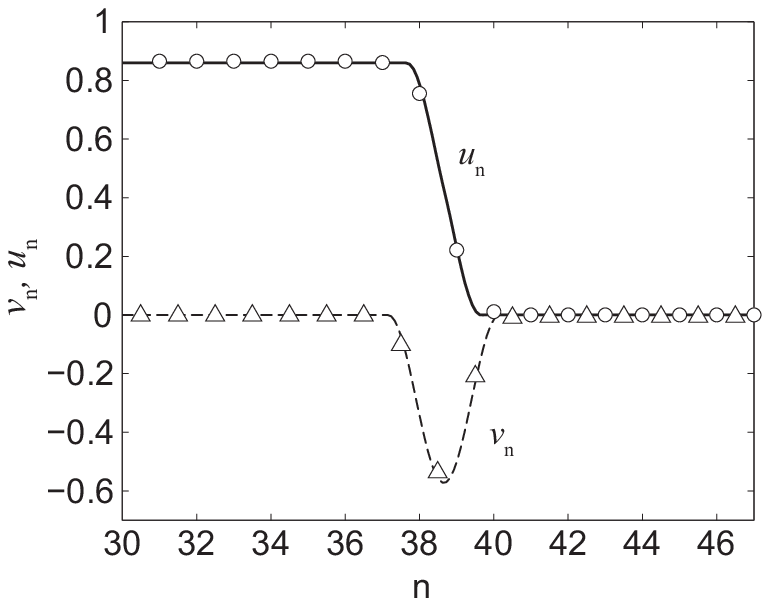}}
\centerline{\includegraphics[width=8cm,clip]{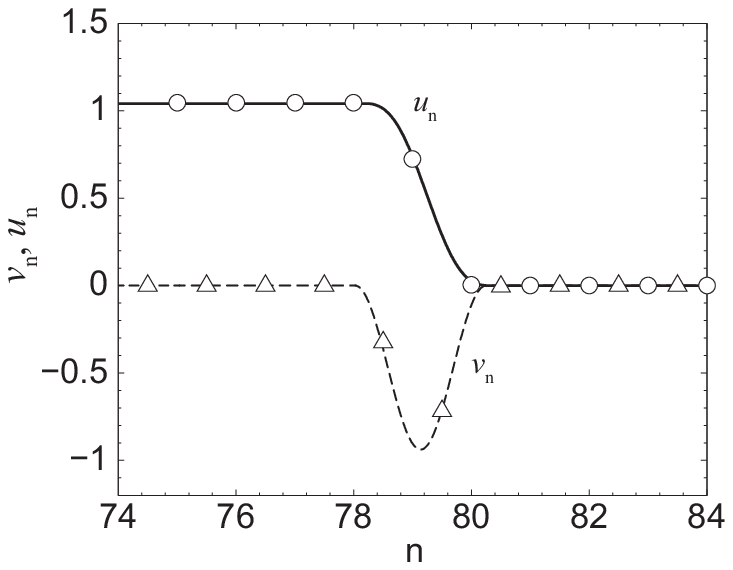}}
    \vskip-3mm\caption{Profile for the two kinks coordinates $u_n$ and
    strains $v_n=u_n-u_{n-1}$. Circles and triangles are numerical results
    and the continuous lines are obtained from Eq.~(\ref{eq:vn})
    with {\em magic} $wave$ $number$ $q\simeq 2\pi/3$ (top),
    when basically two particles are moving at a given
    time,  and
    with $q\simeq\pi$ (bottom) for the amplitude $A$ close to 1, when
    basically one particle is moving at a given time.
    Scaled units are equal to the lattice unit.}
    \clabel{fig:profiles}
    \end{center}
\end{figure}

For the dimensionless wave number $q=2\pi/\lambda$, with $\lambda$
an integer, Eq.~(\ref{eq:vn}) represents a solution where
basically $\lambda$ bonds and $\lambda-1$ particles (the kink
core) are in motion, while the others remain at rest.

We will use the term {\em active} to describe related states of
the different magnitudes. The active particles or coordinates at a
given time or time interval (or phase or phase interval) are those
for which $u_n$ is changing, the active bonds are those for which
$v_n\neq 0$, i.e., they are the compressed ones. For a particle,
the time interval is active when it is moving, and for a bond,
when it is compressed.

If the dimensionless wavelength $\lambda$ is between two integers $m_1$ and $m_2$, the
number of active bonds oscillates between $m_1$ and $m_2$ and the
number of active oscillating particles is between $m_1-1$ and $m_2-1$.

Of particular interest in this work will be $\lambda=3$, with
$q=2\pi/3$, which is called the {\em magic wave
number}~\cite{kosevich04}, and $\lambda=2$, with $q=\pi$, which
will be referred to as the $\pi$-mode. These two values are
extreme cases of localization, with $q=\pi$ being the limit when
only one particle is moving at a given time.

\subsection*{Rotating wave approximation}

The average velocity of the kink can be analytically obtained with the use of the
rotating wave approximation (RWA) as derived in
Ref.~\cite{archilla2013}:

\begin{eqnarray}
V=\frac{\omega}{q}=\frac{1}{(1-A)^{3/4}}\,\cs\,
\frac{\sin(q/2)}{(q/2)} \, .\clabel{eq:RWA}
\end{eqnarray}

Kinks are therefore supersonic. In the small amplitude ($A\rightarrow
0$) and long wavelength ($q\rightarrow 0$) limit,  $V$ tends to the
sound speed $\cs $.

\begin{figure}[h] \begin{center}
\centerline{\includegraphics[width=8cm]{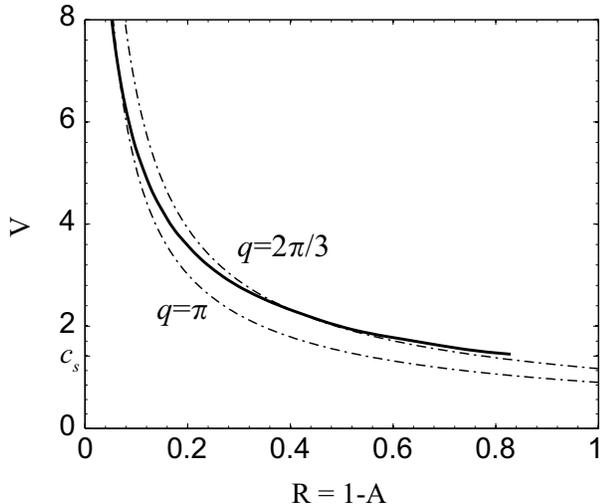}} 
    \vskip-3mm\caption{Kink velocity versus minimal interparticle distance
$R=1-A$, calculated numerically (thick line)
    and analytically from Eq.(\ref{eq:RWA}) with wave number $q=2\pi/3$ (upper dotted-dashed line)
    and $q=\pi$ (lower dotted-dashed line). Scaled units are the lattice constant for distances
     and 2.6~km/s for velocities.}
    \clabel{fig:VAc1RWA}
    \end{center}
\end{figure}
In Ref.~\cite{archilla2013} it was shown that the waveform with the $magic$ $wave$
$number$ demonstrates a good agreement between the {\em ansatz} and
simulations. However this agreement fails at the kink amplitudes
$A$ approaching unity, when the minimal interparticle distance in
the kink (in lattice units) $R=1-A$ diminishes and the $q=\pi$
brings about a much better fit as can be seen in
Fig.~\ref{fig:VAc1RWA}. The conclusion is that the $magic$ $wave$
$number$ is a suitable approximation for an intermediate range of
amplitudes in the Coulomb lattice, in contrast to the FPU chain
where it is valid for all amplitudes\cite{kosevich93,kosevich04},
and that $q$ actually changes continuously with the amplitude in
the Coulomb lattice. The complementary approach is to use
Eq.~(\ref{eq:RWA}) to find the values of the wave number $q$, with
respect to the amplitude $A$ or the velocity $V$. This will be
shown in Fig.~\ref{fig:wavenumber}, where it can be appreciated
that the waveform with the $magic$ $wave$ $number$ is a proper solution in the Coulomb
lattice for the intermediate values of the kink velocity.

\subsection*{Triangular waveform}

For the higher amplitudes, with $A$ close to 1, the waveform
deviates from the sinusoidal one given by  Eq.~(\ref{eq:vn}) and
approaches instead the triangular waveform, which is shown in
Fig.~\ref{fig:triangle}. The almost straight lines in such
waveform mean that the velocity is almost constant except during a
very short interaction time. The system behavior is very similar
to that of the hard-spheres model.
\begin{figure}[h] \begin{center}
\centerline{\includegraphics[width=8cm,clip]{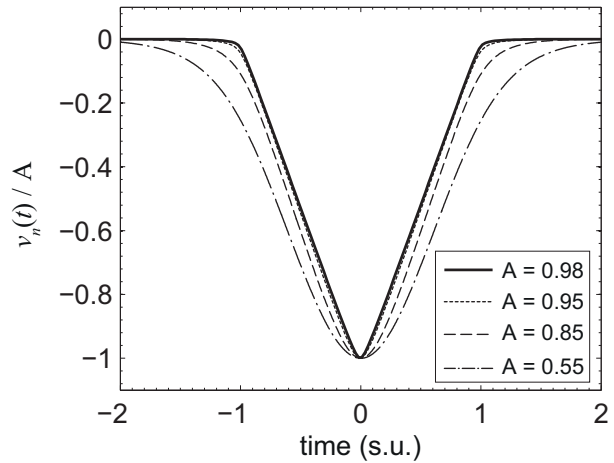}}
\vskip-3mm\caption{Temporal variation of the strain
$v_n=u_n-u_{n-1}$ for different amplitudes. As the amplitude $A$
increases, the shape becomes almost triangular, corresponding to
almost hard-spheres collisions. The amplitude $A$ is given in
lattice units and the scaled time unit is 2~ps. }
    \clabel{fig:triangle}
    \end{center}
\end{figure}

This waveform is also another way of looking at the mode $q=\pi$.
In this mode only one particle is moving at a given time, which
means that there are no forces on the particle acting from its neighbors.
Strictly speaking, the mode $q=\pi$ and an exact triangular
waveform are unattainable because of  the electrostatic Coulomb
forces acting on the particles. However, if the energy of the
particle is large with respect to the change in the potential
during a large part of the path between collisions, the particle
will move almost free during most of the time between collisions.
The triangular waveform has been described as the high-energy
limit for the Lennard-Jones interatomic potential
~\cite{friesecke2002} and has been  observed experimentally in a
system of repelling magnets~\cite{moleron2014}. It is worth
mentioning that the triangular waveform can also be related with
strongly-stretched bonds in the high-energy limit in the
potentials allowing for bond dissociation,  like the Lennard-Jones
potential. The strongly-stretched bonds result in finite (normal)
thermal conductivity in one-dimensional systems with such
interatomic potentials \cite{savin-kosevich2014}.

\subsection{Analytical results}
\subsubsection{Sinusoidal waveform and mode with $q=2\pi/3$}

Some analytical results can be obtained in this model; see also
Ref.~\cite{kosevich04}. Let us consider the wavenumber $q=2\pi/3$
and choose $t=0$ for the time for which  $\phi_n=q\,n-\w t=0$ at
$n=0$ after a change of the origin of $n$ and $t$, but keep the
notation $n$ for generality. If we consider  the time interval
$\Delta t$: $-T/6\leq t<T/6$, there are three active strains:
$v_{n+1}$, $v_n$ and $v_{n-1}$. At the end of the interval, that
is at $t=T/6$, $v_{n-1}$ becomes zero and $v_{n+2}$ starts being
perturbed, so all the indexes $n$ of the particles will change in
a unity.

During $\Delta t$, $u_{n'}=0$ for $n'\geq n+1$, $u_n$ and
$u_{n+1}$ are changing but $u_{n-1}$ has already attained the
value $3A/2$, its final constant value as can be checked by direct
sum. Also, $u_{n'}=3\,A/2$ for $n'< n-1$, their final  value after
the kink has passed over them as seen in Fig.~\ref{fig:profiles}.
The active coordinates are therefore $u_n=-v_{n+1}$ and
$u_{n-1}=-v_n+v_{n+1}$. After some algebra we get
\begin{eqnarray}u_n&=&\Ah+\Ah \cos(\phi_n+2\pi/3)\nonumber \\
u_{n-1}&=&A-\Ah \cos(\phi_n-2\pi/3)\,. \clabel{eq:unn-1}
\end{eqnarray}

We can obtain immediately the kink kinetic energy as
$K=\frac{1}{2}{\dot u_n}^2+\frac{1}{2}{\dot u_{n-1}}^2$, which
results in
\begin{equation}
K=\frac{\w^2 A^2}{8}\big( 1+\frac{1}{2}\cos(2\phi_n)\big)\,,
\clabel{eq:K}
\end{equation}
with maximum value
\begin{equation}
 K_M=\frac{\pi^2}{12}V^2 A^2 \,. 
\end{equation}

The potential energy, with respect to the equilibrium state, is
given by:
 \begin{equation}
 U=\frac{1}{1+v_{n-1}}+\frac{1}{1+v_{n}}+\frac{1}{1+v_{n+1}}-3\, .
 \end{equation}
By substitution, it can be seen that the maximum potential energy
corresponds to the  bond $n$ at its maximum compression, i.e.,
with $\phi_n=0$, while the bonds $n-1,n+1$ have a phase
difference of $\pm 2\pi/3$ with a strain $u_{n-1,n+1}=-A/2(1+cos(\pm
2\pi/3)=-A/4$. Therefore the maximum energy becomes:
\begin{equation}
U_M=\frac{1}{1-A}+\frac{2}{1-A/4}-3\,. 
\clabel{eq:UMsinpi}
\end{equation}

There is also a  minimum potential energy which corresponds to the
limit, in which  only two bonds are different from zero with phases
$\pm \pi/3$ and strain $v_n=-3A/4$. The minimum energy becomes

\begin{equation}
U_m=\frac{2A}{1-3A/4}-2\,. 
 q=2\pi/3\nonumber) \
 \clabel{eq:Umsinpi}
\end{equation}
The kink has always some compression energy $U_m$ above the ground
state.

\subsubsection{Sinusoidal waveform and mode with $q=\pi$}

The properties of sinusoidal kinks with mode $q=\pi$ are easy to
obtain as there are only two active strain variables and one
coordinate, which during the interval $\Delta_\pi t$,  $0\leq t
< T/2$,  are:
\begin{eqnarray}
v_n&=&-\Ah[1+\cos(q\,n-\w t ]\,,\nonumber\\
v_{n+1}&=&-\Ah [1+\cos(q[n+1]-\w
t)]=\nonumber \\ &\mbox{}&-\Ah [1-\cos(q\,n-\w t)]\,, \nonumber \\
u_n&=&-v_{n+1}~;\quad \dot u_n=\Ah\w\sin(q\,n-\w t).
\end{eqnarray}
The kinetic and potential energies, $K=\frac{1}{2}\dot u_n^2$ and
$U=\dfrac{1}{1+v_n}+\dfrac{1}{1+v_{n+1}}-2=\dfrac{1}{1+v_n}+\dfrac{1}{1-v_{n}}-2$,
 can be
obtained. The maximum kinetic, maximum and minimum potential
energies are given by
\begin{eqnarray}
 K_M=\frac{\pi^2}{8}A^2V^2\;;\;
 U_M=\frac{1}{1-A}-1\;;\;
 U_m=\frac{2}{1-A/2}-2. \nonumber \\
\clabel{eq:KMUMsinpi}
\end{eqnarray}

\subsubsection{Triangular waveform and mode with $q=\pi$}

The potential energy fits very well the numerical values, unlike
the kinetic energy, as can be seen in
Figs.~\ref{fig:potentialtres} and \ref{fig:kinetictres}. This is
because the $\pi$ kink is better described by the triangular form
for large energies. Let us suppose that $t=0$ is the time for
which $v_n=-A$, as it takes half a period to change from $-A$ to
$0$, then $V T/2=2$. Therefore, for the interval $0 \leq t<T/2$
the active variables are:
\begin{eqnarray}
&v_n=-A+AVt; \quad &v_{n+1}=-AVt; \nonumber \\
&u_n=-v_{n+1}=AVt; \quad &\dot u_n=AV.
\end{eqnarray}

The kinetic energy is therefore a constant:
\begin{eqnarray}
K=K_M=\frac{1}{2}A^2V^2\, .
  \clabel{eq:KMUMtripi}
\end{eqnarray}
In the numerical simulations, there are short time intervals when
$K$ changes, separated by a larger interval when $K$ is almost
constant.

The maximum and minimum potential energies are identical to that of  the
 $q=\pi$ sinusoidal waveform given by  Eq.~(\ref{eq:KMUMsinpi}).

\begin{figure}[h]
\includegraphics[width=8cm]{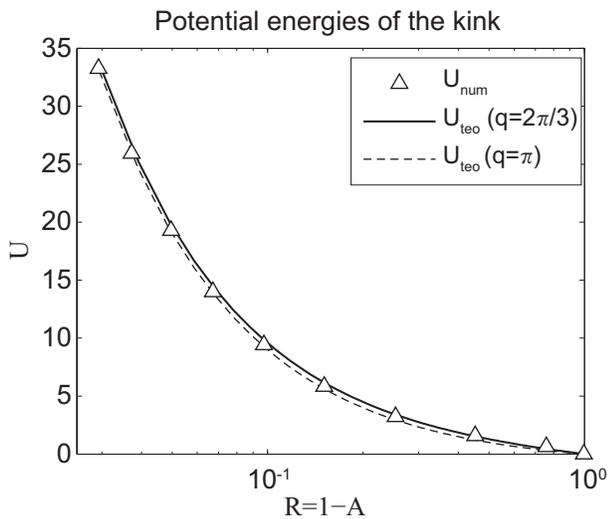}
\caption{Maximum potential energies of the kinks in the Coulomb
potential versus minimal interparticle distance in lattice units
$R=1-A$ (for amplitude $A$ close to 1). The analytical results are
very similar for the waveforms with $q=\pi$ and $q=2\pi/3$, as the
maximum potential energy of the kink depends mainly on the minimal
separation between particles $R$. The values of the dimension
units are the lattice unit and 2.77~eV, respectively.}
\clabel{fig:potentialtres}
\end{figure}
\begin{figure}[h]
\includegraphics[width=8cm]{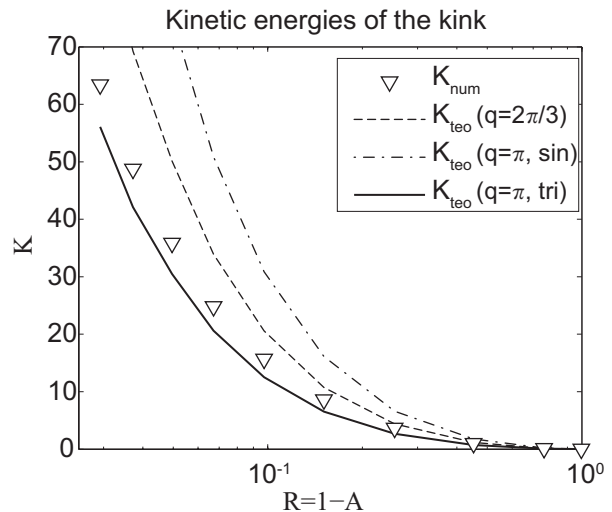}
\caption{Maximum kinetic energies of the kinks in the Coulomb
potential versus minimal interparticle distance in lattice units
$R=1-A$ (for amplitude $A$ close to 1). It can be seen that the
kink with wavenumber $\pi$ is better described by a triangular
waveform than by a sinusoidal one. The values of the dimension
units are the lattice unit and 2.77~eV, respectively.}
\clabel{fig:kinetictres}
\end{figure}

\section{Interaction with several neighbors}
\begin{figure}[h]
\includegraphics[width=8cm]{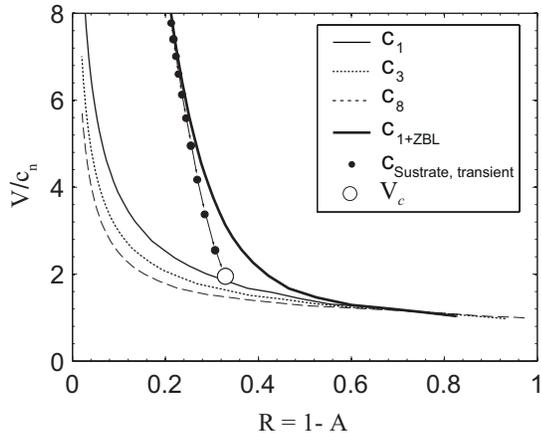}
\caption{Kink velocities as a function of minimal interatomic distance in
lattice units $R=1-A$ ($A$ is kink's amplitude) for several potentials.
Notations in the legend: $C_i$, $i=1,3,8$, refers to the
Coulomb interaction between the first $i$ neighbors. Velocities
are normalized to the sound speed $c_p$ for a system with
interaction between $p$ neighbors, except for the system with a
substrate which is normalized to $c_1=c_s$. It can be seen that the
increase of the number of interacting neighbors slows
the normalized velocity $V/c_p$ of the kink
(but not its absolute velocity). Also, it can be seen that
in the system with a substrate the normalized velocity $V/c_p$ deviates from the
curve $c_{\text{1+ZBL}}$ to some specific velocity $V_c$ on the
$c_1$ curve. The values of the sound velocities are $c_1=c_s=3.7$~km/s, $c_3=5.0$~km/s and $c_8=6.1$~km/s.}
\clabel{fig:VAneighbours}
\end{figure}

\begin{figure}[h]
\includegraphics[width=8cm]{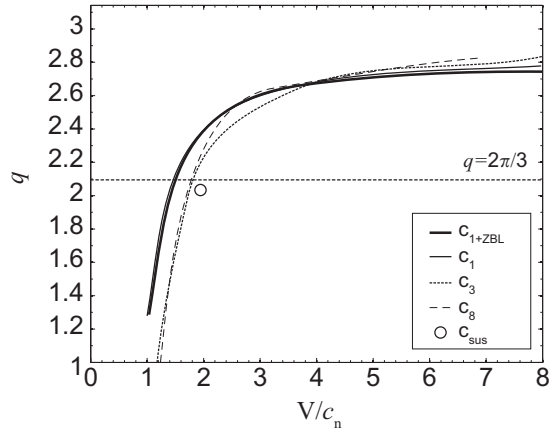}\\
\caption{Best fit for the wavenumber $q$ of the kinks  as a
function of the velocities normalized to the sound speed.
Notations in the legend: $C_i$ is the Coulomb interaction between
the first $i$ neighbors. Note that the {\em magic} $wave$ $number$
$q=2\pi/3$ is also chosen in the system with a substrate. The
values of the sound velocities are $c_1=c_s=3.7$~km/s,
$c_3=5.0$~km/s, and $c_8=6.1$~km/s. } \clabel{fig:wavenumber}
\end{figure}

All the previous results apply to the case of nearest-neighbor
coupling. We have also checked that the kinks still exist and
propagate supersonically when the Coulomb interaction extends
beyond the nearest neighbors. The kinks have similar velocity and
wave number if we take as reference a normalized velocity $V/c_p$,
where $c_p=\cs \big(\sum_{l=1}^p 1/l\,\big)^{1/2}$ is the sound
speed in the Coulomb chain with interactions between $p$ nearest
neighbors. Figure~\ref{fig:VAneighbours} shows the dependence of
the relative velocity $V/c_p$ on the minimum interparticle
distance and Fig.~\ref{fig:wavenumber} shows the dependence of the
best fit for the wave number as a function of the relative
velocity, emphasizing the essential identity of the phenomenon.
Other aspects of these figures will be commented below. For
clarity, only the cases of three and eight neighbors are
represented although up to 30 neighbors have been tested.

It is worth commenting the problem that arises when the
interaction with an infinite number of neighbors is taken into
account. If the pair potential depends on the interparticle
distance $|r|$ as $1/|r|^s$, the long-wave phonon velocity is
finite for $s>1$,  but it diverges with the number of particles
$N$ as $v_\mathrm{ph}\propto \sqrt{\ln(N/2)}$ for the unscreened
Coulomb potential ($s=1$). However, this divergence occurs only in
the electrostatic limit when the electromagnetic wave retardation
is neglected. With an account of the retardation, the long-wave
group velocity tends to the speed of light in vacuum. If the
particles are in a material medium, there is a rearrangement of
the electron density that can be described as a screening of the
Coulomb interaction with some characteristic length
$l_\mathrm{scr}$. The screening  brings about a finite long-wave
phonon velocity
$v_\mathrm{ph}\propto\sqrt{\ln(l_\mathrm{scr}/a)}$, where $a$ is
the lattice constant. As  has been mentioned, we do not study in
depth this problem here and have considered only a few neighbors
for simplicity.

\section{Kinks with short--range ZBL potential}
\clabel{sec:potential}

The minimal interatomic distance in lattice units $R=1-A$,
obtained for fast large-amplitude kinks, is clearly impossible in
realistic systems. At short distances short--range forces appear,
which are produced by the overlapping electronic shells of the two
close atoms. A large number of different repulsive potentials and
screening functions have been proposed over the years, some
determined semi-empirically, others from theoretical calculations.
A much used repulsive potential is the one given by Ziegler,
Biersack and Littmark, the so-called ZBL repulsive potential. It
has been constructed by fitting a universal screening function to
theoretically obtained potentials calculated for a large variety
of atom pairs ~\cite{ziegler2008}. The ZBL potential has the form
\begin{eqnarray}
U_\mathrm{ZBL}(r)=\kc\,\frac{Z_1\,Z_2\,\e^2}{r}\,\phi(\frac{r}{\rho})\,
, \clabel{eq:zbl}
\end{eqnarray}
with $\kc$ being the Coulomb constant, $Z_1$ and $Z_2$ are the
atomic numbers of the involved atoms, and $r$ the distance between
them, $\phi(x)$ is the universal screening function:
\begin{eqnarray}
\phi(x)&=&0.1818\exp(-3.2x)+0.5099\exp(-0.9423x)+\nonumber\\
&&0.2802\exp(-0.4029x)+\nonumber\\
&&0.02817\exp(-0.2016x)~.
\end{eqnarray}
 The screening length is
$\rho=0.8854\,a_0/(Z_1^{0.23}+Z_2^{0.23})$, $a_0=0.529$\AA~ being the
Bohr radius. The ZBL potential describes well the interaction
between neutral atoms. In the case of ions considered here, the
Coulomb potential must also be added, accounting for the repulsion
between the ions. The introduction of the ZBL potential restricts
the interatomic distances to the realistic values. The four terms in
the ZBL potential are important for different range of energies,
but for  $\K$ ions, with energies  up to 200~KeV, which is much
larger that the ones considered here, the interaction potential
can be represented by a single term, which together with the
Coulomb ionic part takes the form,
\begin{eqnarray}
U(r)=\frac{1}{r}+\frac{\alpha}{r}\exp(-\frac{r}{\rho})\, ,
\clabel{eq:potential}
\end{eqnarray}
with $\alpha=184.1$ and $\rho=0.0569$ in dimensionless units,
which in physical units correspond to $2650.6$~eV\AA\,
 and $0.29529$~\AA\,, respectively.  Figure~\ref{fig:potentials}
represents ZBL and Coulomb potentials, with their sum and other
details to be commented on later. Note that around $r\simeq 0.4$,
the combined potential $U(r)$ differs considerably from the
Coulomb potential.

\begin{figure}[h]
\centerline{\includegraphics[width=8cm]{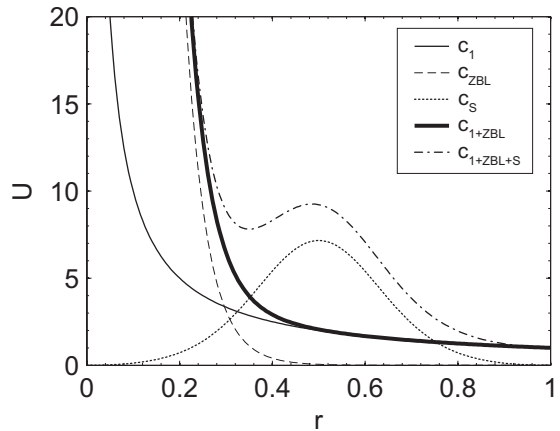}} 
    \vskip-3mm\caption{Interaction potentials $U(r)$ in dimensionless units.
    Coulomb ($c_1$); ZBL ($c_{\text{ZBL}}$); Coulomb+ZBL ($c_{\text{1+ZBL}}$);
    substrate potential ($c_{\text{S}}$) and
    the sum of the Coulomb, ZBL and substrate potentials ($c_{\text{1+ZBL+S}}$).
    The scaled units are 2.77~eV and the lattice unit
    a=5.19~\AA\, for $U$ and $r$, respectively.}
    \clabel{fig:potentials}
 \end{figure}

\begin{figure}[h]
\centerline{\includegraphics[width=8cm]{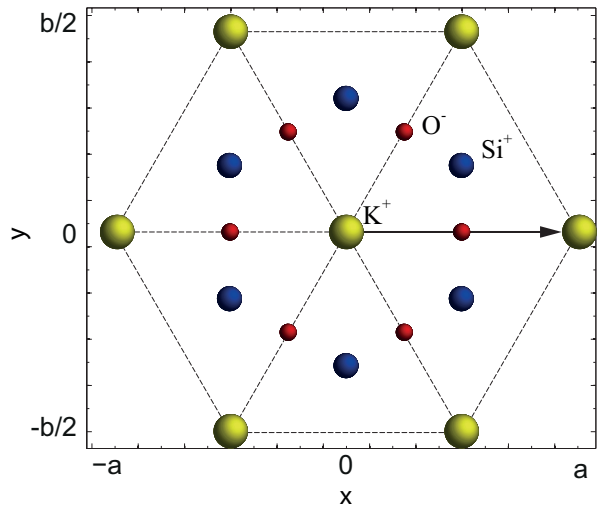}} 
    \vskip-3mm\caption{Projection on the (001) plane of the ions
used in the calculation of the substrate potential. Four planes of
ions are considered, two above and two symmetrical below the \Kt
plane, the closest two with oxygen and the other two with silicon
ions. The path for the central \Kt ion used the in calculation is
shown, note that the O$^{-2}$ ions in the middle of the path are
actually at a distance of 1.68\AA~above and below. The interaction
between the \Kt ions in the central $X$-axis is not taken into
account in constructing the substrate potential as it is taken
into account explicitly. The crystal is continued in the \Kt plane
until the convergence is achieved. Distance between the longer
marks is 1~\AA.}
    \clabel{fig:micahexagonal}
 \end{figure}

The dynamical equations become

\begin{equation}
\ddot u_n=-G_{n+1}+G_{n} -F_{n+1}+F_n\,,
 \clabel{eq:forceFG}
\end{equation}
with $F_n$ given by Eq.~(\ref{eq:ddotvn}) and $G_n$ given by
\begin{equation}
G_n=\frac{\alpha}{1+v_n}\exp(-\frac{1+v_n}{\rho})\Big(\frac{1}{1+v_n}+\frac{1}{\rho}\Big)\,.
\end{equation}

When the joint effect of both the screened Coulomb (ZBL) and bare
Coulomb potentials is considered, i.e., Eq. (\ref{eq:potential}),
numerical simulations show that the behavior of the kinks is not
much different from that observed in the bare Coulomb case,
discussed in the preceding sections. Supersonic kinks propagate
equally well, changing from the $magic$ $wave$ $number$ $q=2\pi/3$
to the proximity of  $q=\pi$ and from the sinusoidal to nearly
triangular waveform when the amplitude $A$ increases.
Figures~\ref{fig:VAneighbours} and \ref{fig:wavenumber} also show
the characteristic curves $V=V(A)$ and $q=q(A)$. The sound
velocity does not change since the ZBL potential is felt only for
very large perturbations. The RWA cannot be obtained analytically
but the numerical RWA fits very well the results of the
simulations.

\section{The effect of the substrate potential: lattice kinks or crowdions}
\clabel{sec:substrate}

In the preceding sections, the interaction with the other atoms in
the crystal was taken into account only implicitly, since the only
effect of the surrounding atoms was to fix the  equilibrium
lattice period and to confine the particles within the crystal. To
better model  the properties of the kinks in a crystal like
muscovite, we take into account explicitly the interaction with
the surrounding atoms in a  simplified mica geometry. The \Kt~ions
occupy the nodes of the hexagonal lattice with a lattice unit
5.19~\AA. There are no other atoms in the \Kt plane, therefore we
need to consider more atoms above and below. We consider two
planes above and two symmetric below.  The closest plane,  at the
distance of 1.6795\AA, is occupied with oxygen ions with charge
--2, their projections on the \Kt plane lie in the middle of the
two nearest-neighbor \Kt~ions. The other layers, at the distance
of 2.2227\AA\, from the \Kt~plane,    are occupied by silicon
ions. They are in the centers of tetrahedra whose three horizontal
vertices are occupied by the oxygen ions. See
Fig.~\ref{fig:micahexagonal} for a sketch. The Si sites are
occupied by Si$^{+4}$ and Al$^{+3}$ ions in the proportion of 3:1,
giving an average charge of +3.75, but we assign them a smaller
charge +2.75 to take into account other atoms in successive
layers, particularly the oxygen ions at the top of the tetrahedra,
and to achieve charge neutrality. We suppose that all the atoms
are in fixed positions except the moving \Kt~ions in a row in the
$[100]$ direction. This is justified by the supersonic speed of
the kinks, that we are interested in, and due to  the weak
interaction between the Si and Al ions compared to the ZBL
interaction between potassium ions.

The interaction between the \Kt~ions, that are in the central
$X$-axis, are not considered in constructing the substrate
potential as it is taken into account explicitly. The lattice is
extended in the (001) \Kt~plane until the convergency of the
potential is achieved.

Specifically the potentials used are the electrostatic
interactions and  the Born-Mayer potentials of the form
$V=A\exp(-r/r_g)$,  given in Ref.~\cite{gedeon2002}. The value
$r_g=0.29$\AA\ is for all the interactions, and the pre-exponential
constants in eV are: $A_{\text{KO}}=3800.125$,
$A_{\text{KSi}}=2762.5$, $A_{\text{OO}}=453.375$,
$A_{\text{0Si}}=1851.25$, and $A_{\text{SiSi}}=1173.125$. For the
\Kt-\Kt~interaction, we use the ZBL potential described above.

We obtain a substrate potential in scaled units ($u_E\simeq
2.77$~eV), which can be described with very good approximation by a
truncated Fourier series:
\begin{eqnarray}
U_s(x)=\sum_{n=0}^4\,U_n\cos(2\pi\,n\,x)\,,
\end{eqnarray}
with coefficients $U_n$ equal to $\{$2.4473, $-3.3490$, 1.0997,
$-0.2302$, 0.0321$\}$. The corresponding linear frequency is
$\w_0=4.48$ or 119~cm$^{-1}$ in physical units, which is close to
110~cm$^{-1}$ determined experimentally~\cite{diaz2000}. Also the
potential well of 20~eV is consistent with molecular dynamics
simulations~\cite{CollinsAM92}. It is represented  in
Fig.~\ref{fig:potentials} together with the other potentials, such
that their relative magnitudes can be compared.

The phonon spectrum becomes an optical one, and the frequency and group
velocity are
\begin{eqnarray}
\w^2&=&\w_0^2+4\,\cs^2\sin^2(q/2)~,\nonumber\\
 V_g&=&\frac{\mathrm{d} \w}{\mathrm{d}
 q}=\frac{\cs^2\sin{q}}{\sqrt{\w_0^2+4\,\cs^2\sin^2(q/2)}}\,.
 \clabel{eq:wqsubs}
\end{eqnarray}
Note that $\cs$ is still the sound speed in the system {\em
without} substrate. The dimensionless phonon frequencies $\w$ are
in the interval between $\w_0=4.48$ and $\w_\mathrm{max}=5.31$.
The group velocity is zero at $q=0$ and $q=\pi$ and reaches its
maximum $V_{\text{g,max}}\simeq 0.4 $ in the proximity of
$q=\pi/2$ with $\lambda\simeq 4$. These features are observed in
the simulations.


\subsection{Qualitative description}

The introduction of the substrate potential does not prevent the
existence of supersonic lattice kinks. The lattice kink, also
called {\em crowdion}, consists of an interstitial atom
propagating very fast in the lattice and leaving behind a vacancy.
The specific feature of the kinks found in the present work is
that its velocity and energy are fixed by the layer$+$substrate
system, let us denote them as $V_c$ and $E_c$ ($c$ for crowdion).
If the initial energy is smaller than $E_c$, the kinks are rapidly
dispersed and disappear, and if it is larger, the excess energy is
radiated as the kink slows down to $V_c$. The specific values in
scaled units are $V_c=2.7387\simeq 2 \cs $ and $E_c=9.4374$,
corresponding to 7.16~km/s and 26.2~eV, respectively. The lattice
kink is supersonic in the two meanings (1)~$V_c>\cs$, where $\cs$
is the sound velocity without substrate and (2)~$V_c$ is much
larger than the maximum phonon group speed $V_{\text{c,max}}\simeq
0.4$ in the system with the substrate, but is not larger than the
maximum phase velocity which is unbounded for $q\rightarrow 0$.

\begin{figure}[htbp]
\centerline{\includegraphics[width=8cm,clip]{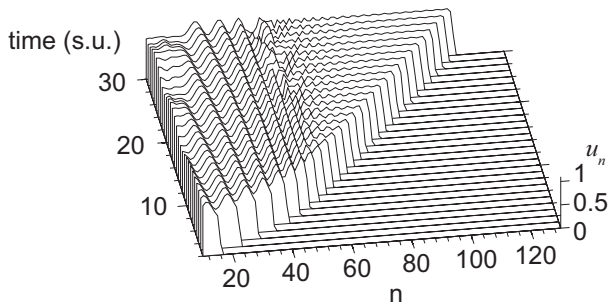}} 
\caption{Representation of the process of dynamical slowing down
of the kink. The nonlinear and linear radiation processes can be
easily distinguished. Nonlinear waves with large amplitude are
first  emitted. Later, the phonons with the wavelength close to
$\lambda\simeq 4$, which corresponds to maximal group velocity,
and with exponentially decaying amplitudes are emitted while the
kink approaches the limit velocity. Scaled units for time and
distance are 0.2~ps and lattice constant 5.19~\AA. }
    \clabel{fig:waterfall}
 \end{figure}

\begin{figure}[htbp]
\centerline{\includegraphics[width=8cm]{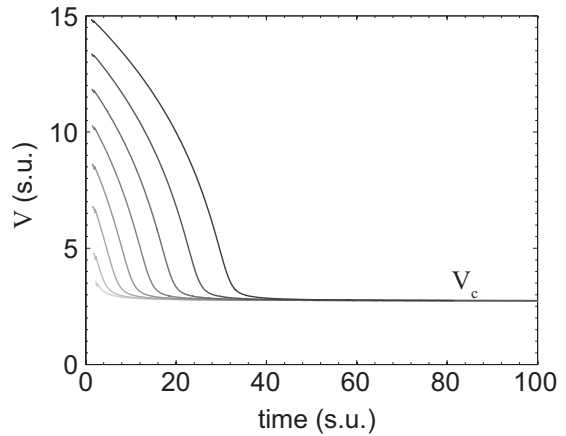}} 
    \vskip-3mm\caption{Velocities of lattice kinks or crowdions versus time in scaled
units for different initial conditions. For velocities $V>V_c$,
the lattice kink slows down until $V=V_c$ is reached. For $V<V_c$,
the kinks are dissipated into low-amplitude phonons. Scaled units
are 0.2~ps and 2.6~km/s.}
    \clabel{fig:Vcrowdion}
 \end{figure}

\subsection{Slowing down processes}

The process of slowing down to $V_c$ is shown in
Figs.~\ref{fig:waterfall} and \ref{fig:Vcrowdion}, where two well
distinguished phases can be identified.

a) {\em Nonlinear radiation: } For an initial energy $E>E_c$, the
kink progressively loses energy. The particles immediately after
the kink perturbation are left with enough energy for nonlinear
vibrations in the potential well bringing about nonlinear wave.
Their frequencies obtained numerically are above the phonon band
with a maximum value of about 6.3. This strong radiative process
is shown in Fig.~\ref{fig:waterfall} for an initial velocity
$V_0=7$. This process is very fast and the loss of energy is
almost linear with time.

b) {\em Linear radiation: } As the lattice kink energy approaches
$E_c$, the amplitude of the tail oscillations and their frequency
decrease, and the kink frequency enters the phonon band, radiating
low amplitude phonons~\cite{BK98,braun2004}. The energy decreases
exponentially with time towards $E_c$.   The wave number of the
radiated phonons can be deduced from the kink speed as each
particle left behind the kink is excited with a delay $\Delta
t=1/V_c$ and, therefore, with a phase difference $q=\w(q)\Delta
t=\w(q)/V_c$. As the phonon wavevector is given by
$q=\w(q)/V_\mathrm{ph}$, where $V_\mathrm{ph}$ is the phase speed
of the phonons, $V_\mathrm{ph}=V_c$. Using the phonon dispersion
relation in Eq.~(\ref{eq:wqsubs}) it is possible to obtain the
phonon wavenumber and wavelength $\lambda _{\text{ph}}=3.5$ which
is the observed one in the simulations. Similar process has been
described in Ref.~\cite{BK98} and in references therein for
subsonic kinks. However, there is an important difference:  The
subsonic kinks, described in those works, radiate continuously and
eventually stop.

From the comparison of Figs. 7 and 12 with Figs. 9 and 13 we
conclude that the amplitude $A=0.67$ of relative particle
displacements of supersonic kink with the unique velocity is
determined by the interatomic distance $R$ at the $minimum$ of the
$total$, interatomic plus substrate, potential, $R=1-A\approx
0.33$ in lattice units. This value of the kink amplitude ensures
the absence of kink oscillations and correspondingly the absence
of phonon radiation into the chain by the supersonic kink. We also
want to emphasize that we do not observe a discrete spectrum (a
set of possible values) of the velocity of the supersonic kink in
our system, even when we start with high initial kink velocity,
see Fig. 12. This is in contrast with the prediction of supersonic
multi-solitons (lattice $N$-solitons) in Frenkel-Kontorova model
with nonlinear interparticle coupling in Refs.
\cite{braun2004,savin95,zolotaryuk97}. We relate this finding with
the extreme discreteness of our kink with the $magic$ $wave$
$number$ $q\simeq 2\pi/3$, which as shown in Fig. 8 corresponds to
only two particles moving at a given time, cf. Ref.
\cite{kosevich04}, and which does not allow for different
matchings with the substrate potential. The uniqueness of the
observed supersonic kink velocity we relate also with possible
dynamical instability of lattice multi-solitons (bound states of
supersonic kinks) in the considered system.

\subsection{Double kink}

\begin{figure}[htbp]
    \centering
    \includegraphics[width=8cm]{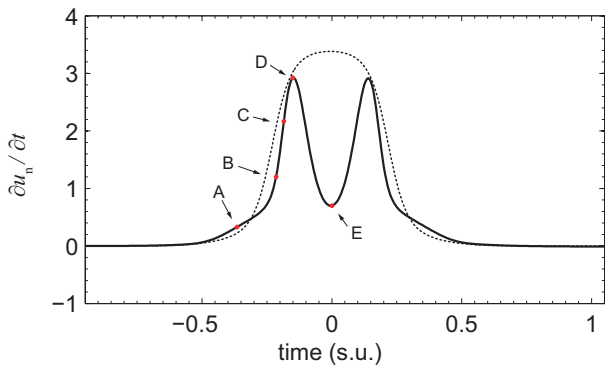}
    \caption{Particle velocity waveform measured at $n=470$ for
the supersonic kink with amplitude  $A=0.67$ and minimal
interatomic distance in lattice units $R=1-A=0.33$ for Coulomb+ZBL
potential with (continuous  line) and  without (dotted line)
substrate potential. The instants when the ZBL interaction acts
(near A) and the minimum velocity at the top of the potential
barrier (E) are easily identified. The configurations at those
times can be seen in Fig.~\ref{fig:stable-kink-PN-states}. Scaled
units are 0.2~ps and 2.6~km/s.}
    \clabel{fig:particlevelocity}
\end{figure}

The particle velocity $\dot u_n$ as a function of time for the
stable kink is represented in Fig.~\ref{fig:particlevelocity}. Due
to the extreme discreteness of the kink, it is not practical to
represent $\dot u_n$ an a function of $n$, while the double-kink
structure of the found kink in the coordinates, shown in
Fig.~(\ref{fig:stable-kink-PN-states})-(a), is evident as a
function of time.


\begin{figure*}[t]
    \centering
    \includegraphics[width=14cm]{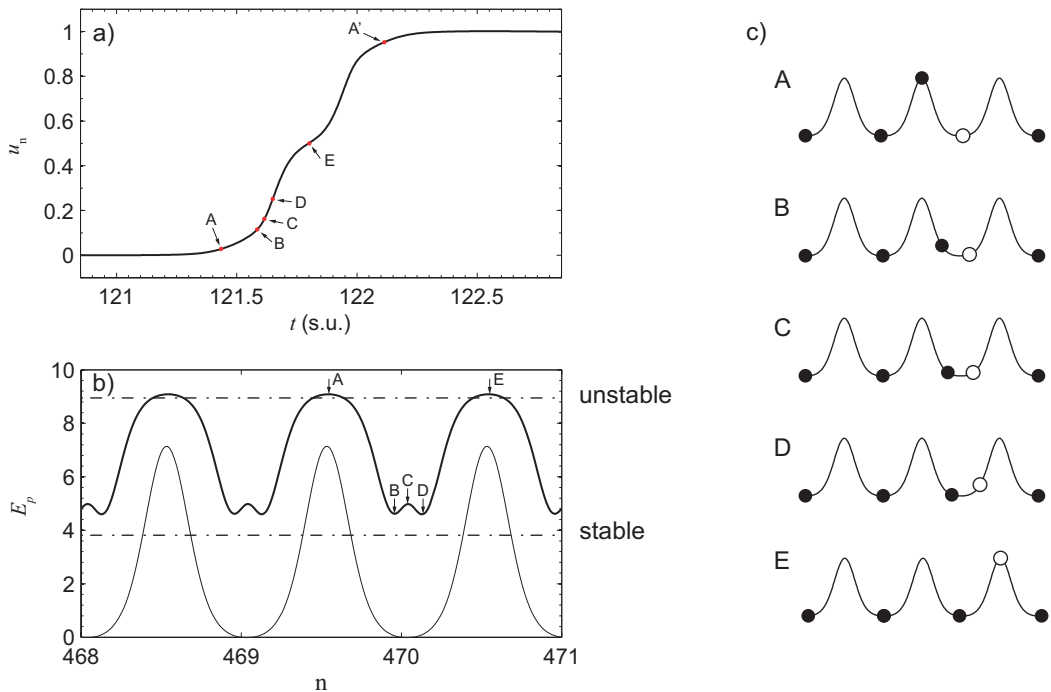}
    \caption{(a) Particle
    displacement waveform for the established lattice kink.
    (b)~Potential energy seen by the kink during propagation (thick
    line) and periodic
substrate  potential (thin line), horizontal lines are the
energies corresponding to the equilibrium stable (C) and unstable
(A) interstitials. (c) Particle  configurations corresponding to
the points $A-E$, shown in (a) and (b). Scaled units are lattice
unit 5.19~\AA~ for distances, 0.2~ps for time and  2.77~eV for energies.}
    \clabel{fig:stable-kink-PN-states}
\end{figure*}

The extreme discreteness of the kink makes possible the detailed
description of the double-kink profile as shown in
Fig.~\ref{fig:stable-kink-PN-states}~(c). The particle $n$,
represented by a white circle, which is initially at rest at the
bottom of a potential well, experiences  two sequential
collisions: one when it was hit by the particle $n-1$ and is
accelerated afterwards, and a second one when it hits the particle
$n+1$ and is decelerated, attaining almost zero velocity at the
bottom of the potential well (the particle velocity is different
from zero during the radiation process). In-between the two
collisions, the particle finds the substrate potential barrier
between the  sites and experiences a decrease in velocity while
going uphill which is followed by an acceleration while going
downhill.

Note that the states A and E in the
Fig.~\ref{fig:stable-kink-PN-states} have exactly the same energy,
but the particle $n$ has moved only half a lattice site. This is
the first kink of the double-kink waveform shown in
Fig.~\ref{fig:stable-kink-PN-states}~(a). The evolution of the
particle $n$, going downhill and hitting the  $n+1$  particle
until it stops, forms the second kink in the double-kink
structure.

This process can also be seen in terms of the kink energy in
Fig.~\ref{fig:stable-kink-PN-states}~(b). Two identical
oscillations of the particle's potential energy $E_p$ happen while
the particle $n$ travels  one lattice site. There is a local
maximum at point $C$, corresponding to the minimum distance
between particles with the short-range ZBL interaction. The
horizontal dashed-dotted lines indicates the energies for the
equilibrium interstitial configurations, with two particles inside
a potential well (stable), or one particle at the top of the
potential barrier (unstable), with the energy difference
corresponding to the Peierls-Nabarro (PN) barrier. The potential
energy is always above the stable interstitial energy as the
lattice has no time to relax, bringing about an {\em adiabatic} PN
barrier. The kink always has finite kinetic energy, with the
minimum reached in the configuration $A$.

\section{Summary}
\clabel{sec:conclusions}

In this paper we have developed a comprehensive dynamical model of
the localized lattice excitations in the low dimensional system
using realistic potentials corresponding to a row of ions in the
silicate layer of the mica muscovite crystal.

Our objective was  to determine what kind of $propagating$ nonlinear localized excitations
can exist in a layered crystal with realistic parameters and with what characteristics.
The choice of the parameters of the mica muscovite is motivated by the
fact that many of the dark tracks that appear in sheets of this
material are consistent with the in-layer propagation of localized vibrational
excitations along the close-packed lattice lines of ions, and an
experiment has demonstrated that localized energetic excitations  can travel along the
close-packed lattice directions, being able to eject an atom at the opposite surface.

The modeling of the system has followed the process of increasing
complexity for better understanding which effect is responsible
for which characteristic of the model. In the starting model used
in  preliminary publications
\cite{archilla2013,archilla2014medium}, only \Kt ions with
nearest-neighbor Coulomb repulsion were taken into account, for
which we have found that very fast supersonic kinks can propagate.
They are extremely localized, with only two particles or, for
higher energies, only a single particle in motion at the same
time.

In the present paper we have performed analytical calculations of
the displacement patterns and energies of the supersonic kinks
within the proposed dynamical model and compare them with
numerical simulations. We have also shown that the introduction of
the interaction with several non-nearest neighbors does not
produce significative changes as long as the sound speed in each
system is taken as a reference. The extremely short minimal
interatomic distances in the kink has motivated us to introduce
more realistic short-range Ziegler-Biersack-Littmark repulsive
potential. In the improved dynamical model, kinks propagate
equally well and with as high energy as desired, and with
realistic minimal interatomic distances.

The next step was the introduction of a periodic substrate
potential using empirical potentials and the geometry of the
layered crystal. Supersonic kinks continue to propagate without
losing energy but with several important properties: (a)~The main
one is that the system selects only single velocity and single
energy of the kink; (b)~The energy of the lattice kink is larger
than the one which is needed for atom ejection at the surface, and
is smaller than the one of the proposed sources of energy, the
recoil of a \Kt ion due to $\beta$~emission; and (c) The found
kinks can be described as double-kinks or bi-solitons depending on
the dynamical variable chosen. The unique average velocity of the
supersonic kink on the periodic substrate potential we relate with
the kink amplitude of the relative particle displacements which is
determined by the interatomic distance corresponding to the
$minimum$ of the $total$, interparticle plus substrate, lattice
potential.

The found kinks are ultra discrete and can be described with the
$magic$ $wave$ $number$ $q\simeq 2\pi/3$, which was previously
revealed in the nonlinear sinusoidal waves and supersonic kinks in
the Fermi-Pasta-Ulam lattice \cite{kosevich93,kosevich04}.  The
extreme discreteness of the supersonic kinks observed in our work,
with basically two particles moving at the same time, allows for
the detailed interpretation of their double-kink structure which
is not possible for the multi-kinks without an account for the
lattice discreteness. The double-kink structure is produced by the
matching of the two sequential collisions experienced by a
particle with the process of going over the substrate potential
barrier between neighboring sites.

The computed energy of the supersonic kinks found in the realistic
lattice potential considered is approximately 26~eV. Such energy
can be provided by the recoil of isotopes of potassium after
radioactive decay and it is larger than the sputtering energy.
This value of the characteristic energy of the supersonic kinks
found allows us to assume that  the tracks found in mica muscovite
crystals can be related with the propagating lattice kinks.

\acknowledgments JFRA, VSM and LMGR acknowledge financial support
from MICINN,  projects FIS2008-04848, FIS2011-29731-C02-02 and
MTM2012-36740-C02-02. All authors acknowledge Prof. F. M. Russell
for ongoing discussions.


\begin{thebibliography}{10}

\bibitem{durrani2001}
S.~A. Durrani.
\newblock Nuclear tracks: A success story of the 20th century.
\newblock {\em Rad. Meas}, {{\bf 34}} (2001) 5--13.
\bibitem{durrani2008}
S.~A. Durrani.
\newblock Nuclear tracks today: Strengths, weaknesses, challenges.
\newblock {\em Rad. Meas}, {{\bf 43}} (2008) S26--S33.
\bibitem{silkbarnes59}
E.~C.~H. Silk and R.~S. Barnes.
\newblock Examination of fission fragment tracks with an electron microscope.
\newblock {\em Phylos. Mag.}, {{\bf 4}} (1959) 970--971.
\bibitem{pricewalker62}
P.~B. Price and R.~M. Walker.
\newblock Observation of fossil particle tracks in natural micas.
\newblock {\em Nature}, {{\bf 196}} (1962) 732--734.
\bibitem{fleischer2011}
R.~Fleischer.
\newblock {\em Tracks to Innovation. Nuclear Tracks in Science and Technology}.
\newblock Springer, New York, 2011.
\bibitem{snowdenifft1995}
D.~P. {Snowden-Ifft}, E.~S. Freeman and P.~B. Price.
\newblock Limits on dark-matter using ancient mica.
\newblock {\em Phys. Rev. Lett.}, {{\bf 74}}, 21 (1995) 4133--4136.
\bibitem{AmMin01a}
M.~D. Alba, A.~I. Becerro, M.~A. Castro and A.~C. Perdig\'{o}n.
\newblock Hydrothermal reactivity of {L}u-saturated smectites: Part {I}. a
  long-range order study.
\newblock {\em Am. Mineral.}, {{\bf 86}} (2001) 115.
\bibitem{HYISNM02}
Z.~L. Hong, H.~Yoshida, Y.~Ikuhara, T.~Sakuma, T.~Nishimura and
M.~Mitomo.
\newblock The effect of additive on sintering behavior and strength retention
  in silicon nitride with {RE}-disilicate.
\newblock {\em J. Eur. Ceram. Soc.}, {{\bf 22}} (2002) 527.
\bibitem{ACANT06}
J.~F.~R. Archilla, J.~Cuevas, M.~D. Alba, M.~Naranjo and J.~M.
Trillo.
\newblock Discrete breathers for understanding reconstructive mineral processes
  at low temperatures.
\newblock {\em J. Phys. Chem. B}, {{\bf 110}}, 47 (2006) 24112--24120.
\bibitem{DSA11}
V.~I. Dubinko, P.~A. Selyshchev and J.~F.~R. Archilla.
\newblock Reaction-rate theory with account of the crystal anharmonicity.
\newblock {\em Phys. Rev. E}, {{\bf 83}} (2011) 041124.
\bibitem{russell67b}
F.~M. Russell.
\newblock Tracks in mica caused by electron showers.
\newblock {\em Nature}, {{\bf 216}} (1967) 907--909.
\bibitem{russell68}
F.~M. Russell.
\newblock Duration of sensitive period for track recording in mica.
\newblock {\em Nature}, {{\bf 217}} (1967) 51--52.
\bibitem{russell88a}
F.~M. Russell.
\newblock Positive charge transport in layered crystalline solids.
\newblock {\em Phys. Lett. A}, {{\bf 130}} (1988) 489--491.
\bibitem{russell88b}
F.~Russell.
\newblock Identification and selection criteria for charged lepton tracks in
  mica.
\newblock {\em Nucl. Tracks. Rad. Meas.}, {{\bf 15}} (1988) 41--44.
\bibitem{NDSA40}
J.~Cameron and B.~Singh.
\newblock Nuclear data sheets for {A}=40.
\newblock {\em Nucl. Data Sheets}, {{\bf 102}}, {2} (2004) 293--513.
\bibitem{radionuclides2012}
{X. Mougeot and R. G. Helmer}.
\newblock {LNE-LNHB/CEA}--{T}able de {R}adion\'uclides, {K}-40 tables.
\newblock {\tt http://www.nucleide.org }, 2012.
\bibitem{schlosserrussell94}
D.~Schl\"{o}{\ss}er, K.~Kroneberger, M.~Schosnig, F.~M. Russell
and K.~O.
  Groeneveld.
\newblock Search for solitons in solids.
\newblock {\em Rad. Meas}, {{\bf 23}} (1994) 209--213.
\bibitem{russelleilbeck07}
F.~M. Russell and J.~C. Eilbeck.
\newblock Evidence for moving breathers in a layered crystal insulator at
  300{K}.
\newblock {\em Europhys. Lett.}, {{\bf 78}} (2007) 10004.
\bibitem{kudriavtsev2005}
Y.~Kudriavtsev, A.~Villegas, A.~Godines and R.~Asomoza.
\newblock Calculation of the surface binding energy for ion sputtered
  particles.
\newblock {\em Appl. Surf. Sci.}, {{\bf 239}}, 3-4 (2005) 273--278.
\bibitem{Dou11}
Q.~Dou, J.~Cuevas, J.~C. Eilbeck and F.~M. Russell.
\newblock Breathers and kinks in a simulated crystal experiment.
\newblock {\em Discret. Contin. Dyn. Syst. S}, {{\bf 4}} (2011) 1107--1118.
\bibitem{archilla2013}
J.~F.~R. Archilla, Y.~Kosevich, N.~Jim\'enez, V.~J.
S\'{a}nchez-Morcillo and
  L.~M. Garc\'{\i}a-Raffi.
\newblock Moving excitations in cation lattices.
\newblock {\em Ukr. J. Phys.}, {{\bf 58}}, 7 (2013) 646--656.
\bibitem{archilla2014medium}
J.~F.~R. Archilla, Y.~A. Kosevich, N.~Jim\'enez, V.~J.
S\'anchez-Morcillo and
  L.~M. Garc\'{i}a-Raffi.
\newblock Supersonic kinks in {C}oulomb lattices.
\newblock In R.~Carretero-Gonz\'alez et~al., editors, {\em Localized
  Excitations in Nonlinear Complex Systems}, pages 317--331. Springer, New
  York, 2014.
\bibitem{ziegler2008}
J.~Biersack, J.~Ziegler and M.~Ziegler.
\newblock {\em {SRIM} - The Stopping and Range of Ions in Matter}.
\newblock Published by J.P. Ziegler, Chester, Maryland, 2008.
\bibitem{savin-kosevich2014}
A.~V. Savin and Y.~A. Kosevich.
\newblock Thermal conductivity of molecular chains with asymmetric potentials
  of pair interactions.
\newblock {\em Phys. Rev. E}, {{\bf 89}} (2014) 032102.
\bibitem{gibson1960}
J.~B. Gibson, A.~N. Goland, M.~Milgram and G.~H. Vineyard.
\newblock Dynamics of radiation damage.
\newblock {\em Phys. Rev.}, {{\bf 120}} (1960) 1229--1253.
\bibitem{meftah1994}
A.~Meftah, F.~Brisard, J.~M. Costantini, E.~Dooryhee, M.~Hage-Ali,
M.~Hervieu,
  J.~P. Stoquert, F.~Studer and M.~Toulemonde.
\newblock Track formation in {SiO}$_2$ quartz and the thermal-spike mechanism.
\newblock {\em Phys. Rev. B}, {{\bf 49}} (1994) 12457--12463.
\bibitem{trautmann2000}
C.~Trautmann, S.~Klaum\"unzer and H.~Trinkaus.
\newblock Effect of stress on track formation in amorphous iron boron alloy:
  Ion tracks as elastic inclusions.
\newblock {\em Phys. Rev. Lett.}, {{\bf 85}} (2000) 3648--3651.
\bibitem{FKM}
Y.~Frenkel and T.~Kontorova.
\newblock On the theory of plastic deformation and twinning.
\newblock {\em Phys. Z. Sowjetunion}, {{\bf 13}} (1938) 1--10.
\bibitem{chaikin95}
P.~M. Chaikin and T.~C. Lubensky.
\newblock {\em Principles of Condensed Matter Physics}.
\newblock Cambridge University Press, Cambridge, 1995.
\bibitem{BK98}
O.~M. Braun and Y.~S. Kivshar.
\newblock Nonlinear dynamics of the {F}renkel--{K}ontorova model.
\newblock {\em Phys. Rep.}, {{\bf 306}} (1998) 1--108.
\bibitem{braun2004}
Y.~K. O.M.~Braun.
\newblock {\em The Frenkel-Kontorova Model}.
\newblock Springer, Berlin, 2004.
\bibitem{kosevich73}
A.~Kosevich and A.~Kovalev.
\newblock The supersonic motion of a crowdion. {T}he one dimensional model with
  nonlinear interaction between the nearest neighbors.
\newblock {\em Solid State Commun.}, {{\bf 12}} (1973) 763--764.
\bibitem{milchev90}
A.~Milchev.
\newblock Breakup threshold of solitons in systems with nonconvex interactions.
\newblock {\em Phys. Rev. B}, {{\bf 42}} (1990) 6727--6729.
\bibitem{savin95}
A.~Savin.
\newblock Supersonic regimes of motion of a topological soliton.
\newblock {\em Sov. Phys. JETP}, {{\bf 81}}, 3 (1995) 608--613.
\bibitem{zolotaryuk97}
Y.~Zolotaryuk, J.~Eilbeck and A.~Savin.
\newblock Bound states of lattice solitons and their bifurcations.
\newblock {\em Physica D}, {{\bf 108}} (1997) 81--91.
\bibitem{ni-kosevich2014}
Y.~Ni, Y.~A. Kosevich, S.~Xiong, Y.~Chalopin and S.~Volz.
\newblock Substrate-induced cross-plane thermal propagative modes in few-layer
  graphene.
\newblock {\em Phys. Rev. B}, {{\bf 89}} (2014) 205413.
\bibitem{pomeau1986}
Y.~Pomeau.
\newblock Front motion, metastability and subcritical bifurcations in
  hydrodynamics.
\newblock {\em Physica D}, {{\bf 23}}, 1-3 (1986) 3 -- 11.
\bibitem{cross1999}
M.~C. Cross and P.~C. Hohenberg.
\newblock Pattern formation outside of equilibrium.
\newblock {\em Rev. Mod. Phys.}, {{\bf 65}} (1993) 851--1112.
\bibitem{clerc2011}
M.~G. Clerc, R.~G. El\'{\i}as and R.~G. Rojas.
\newblock Continuous description of lattice discreteness effects in front
  propagation.
\newblock {\em Phil. Trans. R. Soc. A}, {{\bf 369}}, 1935 (2011) 412--424.
\bibitem{matthews2011}
P.~Matthews and H.~Susanto.
\newblock Variational approximations to homoclinic snaking in continuous and
  discrete systems.
\newblock {\em Phys. Rev. E}, {{\bf 84}} (2011) 066207.
\bibitem{kosevich93}
Y.~A. Kosevich.
\newblock Nonlinear sinusoidal waves and their superposition in anharmonic
  lattices.
\newblock {\em Phys. Rev. Lett.}, {{\bf 71}} (1993) 2058--2061.
\bibitem{kosevich04}
Y.~A. Kosevich, R.~Khomeriki and S.~Ruffo.
\newblock Supersonic discrete kink-solitons and sinusoidal patterns with {\em
  magic} wave number in anharmonic lattices.
\newblock {\em Europhys. Lett.}, {{\bf 66}} (2004) 21--27.
\bibitem{poggi1997}
P.~Poggi and S.~Ruffo.
\newblock Exact solutions in the {FPU} oscillator chain.
\newblock {\em Physica D}, {{\bf 103}}, 1-4 (1997) 251 -- 272.
\bibitem{friesecke2002}
G.~Friesecke and K.~Matthies.
\newblock Atomic-scale localization of high-energy solitary waves on lattices.
\newblock {\em Physica D}, {{\bf 171}}, 4 (2002) 211 -- 220.
\bibitem{moleron2014}
M.~Moler\'on, A.~Leonard and C.~Daraio.
\newblock Solitary waves in a chain of repelling magnets.
\newblock {\em J. Appl. Phys.}, {{\bf 115}}, 18 (2014) 184901.
\bibitem{gedeon2002}
O.~Gedeon, J.~Machacek and M.~Liska.
\newblock Static energy hypersurface mapping of potassium cations in potassium
  silicate glasses.
\newblock {\em Phys. Chem. Glass.}, {{\bf 43}}, 5 (2002) 241--246.
\bibitem{diaz2000}
M.~Diaz, V.~C. Farmer and R.~Prost.
\newblock Characterization and assignment of far infrared absorption bands of
  {K}$^+$ in muscovite.
\newblock {\em Clays Clay Miner.}, {{\bf 48}} (2000) 433--438.
\bibitem{CollinsAM92}
D.~R. Collins and C.~R.~A. Catlow.
\newblock Computer simulation of structure and cohesive properties of micas.
\newblock {\em Am. Mineral.}, {{\bf 77}}, 11-12 (1992) 1172--1181.
\end{thebibliography}

\end{document}